\documentclass[twocolumn,epjc3]{svjour3}  

\RequirePackage[T1]{fontenc}

\smartqed  

\RequirePackage{graphicx}
\RequirePackage{mathptmx}      
\RequirePackage{flushend}
\RequirePackage[numbers,sort&compress]{natbib}
\RequirePackage[colorlinks,citecolor=blue,urlcolor=blue,linkcolor=blue]{hyperref}

\usepackage{graphics}
\usepackage{siunitx}
\usepackage{soul}
\usepackage[dvipsnames]{xcolor}

\newcommand\ee{e$^+$e$^-$}
\newcommand\thetaee{$\Theta_{ee}$}

\usepackage[numbers]{natbib}

\journalname{Eur. Phys. J. C}

\usepackage{lineno}

\begin{document}

\title{Search for the X17  particle in $^{7}\mathrm{Li}(\mathrm{p},\mathrm{e}^+ \mathrm{e}^{-}) ^{8}\mathrm{Be}$  processes  with the MEG II detector.}





\newcommand*{\INFNPi}{INFN Sezione di Pisa$^{a}$; Dipartimento di Fisica$^{b}$ dell'Universit\`a, Largo B.~Pontecorvo~3, 56127 Pisa, Italy}
\newcommand*{\INFNGe}{INFN Sezione di Genova$^{a}$; Dipartimento di Fisica$^{b}$ dell'Universit\`a, Via Dodecaneso 33, 16146 Genova, Italy}
\newcommand*{\INFNPv}{INFN Sezione di Pavia$^{a}$; Dipartimento di Fisica$^{b}$ dell'Universit\`a, Via Bassi 6, 27100 Pavia, Italy}
\newcommand*{\INFNRm}{INFN Sezione di Roma$^{a}$; Dipartimento di Fisica$^{b}$ dell'Universit\`a ``Sapienza'', Piazzale A.~Moro, 00185 Roma, Italy}
\newcommand*{\INFNNa}{INFN Sezione di Napoli, Via Cintia, 80126 Napoli, Italy}
\newcommand*{\INFNLe}{INFN Sezione di Lecce$^{a}$; Dipartimento di Matematica e Fisica$^{b}$ dell'Universit\`a del Salento, Via per Arnesano, 73100 Lecce, Italy}
\newcommand*{\ICEPP} {ICEPP, The University of Tokyo, 7-3-1 Hongo, Bunkyo-ku, Tokyo 113-0033, Japan }
\newcommand*{\Kobe} {Kobe University, 1-1 Rokkodai-cho, Nada-ku, Kobe, Hyogo 657-8501, Japan}
\newcommand*{\UCI}   {University of California, Irvine, CA 92697, USA}
\newcommand*{\KEK}   {KEK, High Energy Accelerator Research Organization, 1-1 Oho, Tsukuba, Ibaraki 305-0801, Japan}
\newcommand*{\PSI}   {Paul Scherrer Institut PSI, 5232 Villigen, Switzerland}
\newcommand*{\Waseda}{Research Institute for Science and Engineering, Waseda~University, 3-4-1 Okubo, Shinjuku-ku, Tokyo 169-8555, Japan}
\newcommand*{\BINP}  {Budker Institute of Nuclear Physics of Siberian Branch of Russian Academy of Sciences, 630090 Novosibirsk, Russia}
\newcommand*{\JINR}  {Joint Institute for Nuclear Research, 141980 Dubna, Russia}
\newcommand*{\ETHZ}  {Institute for Particle Physics and Astrophysics, ETH Z\" urich, 
Otto-Stern-Weg 5, 8093 Z\" urich, Switzerland}
\newcommand*{\NOVS}  {Novosibirsk State University, 630090 Novosibirsk, Russia}
\newcommand*{\NOVST} {Novosibirsk State Technical University, 630092 Novosibirsk, Russia}
\newcommand*{\ScuolaPi}{Scuola Normale Superiore, Piazza dei Cavalieri 7, 56126 Pisa, Italy}
\newcommand*{\INFNLNF}{\textit{Present Address}: INFN, Laboratori Nazionali di Frascati, Via 
E. Fermi, 40-00044 Frascati, Rome, Italy}
\newcommand*{\Liverpool}{Oliver Lodge Laboratory, University of Liverpool, Liverpool, L69 7ZE, United Kingdom}

\date{Received: date / Accepted: date}

\author{
The MEG~II collaboration\\\\
        K.~Afanaciev~\thanksref{addr12} \and
        A.~M.~Baldini~\thanksref{addr1}$^{a}$ \and
        S.~Ban~\thanksref{addr10} \and
        H.~Benmansour~\thanksref{addr1}$^{ab}$\thanksref{e1} \and
        G.~Boca~\thanksref{addr4}$^{ab}$ \and         P.~W.~Cattaneo~\thanksref{addr4}$^{a}$ \and
        G.~Cavoto~\thanksref{addr5}$^{ab}$\thanksref{e1}\and
        F.~Cei~\thanksref{addr1}$^{ab}$ \and
        M.~Chiappini~\thanksref{addr1}$^{ab}$ \and
        A.~Corvaglia~\thanksref{addr6}$^{a}$ \and
        G.~Dal~Maso\thanksref{addr2,addr16} \and
        A.~De~Bari~\thanksref{addr4}$^{a}$ \and
        M.~De~Gerone~\thanksref{addr3}$^{a}$ \and
        L.~Ferrari~Barusso~\thanksref{addr3}$^{ab}$ \and
        M.~Francesconi~\thanksref{addr17} \and 
        L.~Galli~\thanksref{addr1}$^{a}$ \and
        G.~Gallucci~\thanksref{addr1}$^{a}$ \and
        F.~Gatti~\thanksref{addr3}$^{ab}$ \and
        L.~Gerritzen~\thanksref{addr10}  \and
        F.~Grancagnolo~\thanksref{addr6}$^{a}$ \and
        E.~G.~Grandoni~\thanksref{addr1}$^{ab}$ \and 
        M.~Grassi~\thanksref{addr1}$^{a}$ \and 
        D.~N.~Grigoriev~\thanksref{addr7,addr8,addr9} \and
        M.~Hildebrandt~\thanksref{addr2} \and
        F.~Ignatov~\thanksref{addr15} \and
        F.~Ikeda~\thanksref{addr10}  \and
        T.~Iwamoto~\thanksref{addr10}  \and
        S.~Karpov~\thanksref{addr7,addr9} \and
        P.-R.~Kettle~\thanksref{addr2} \and
        N.~Khomutov~\thanksref{addr12} \and
        A.~Kolesnikov~\thanksref{addr12}  \and
        N.~Kravchuk~\thanksref{addr12}  \and
        V.~Krylov~\thanksref{addr12} \and
        N.~Kuchinskiy~\thanksref{addr12}  \and
        F. Leonetti~\thanksref{addr1}$^{ab}$ \and
        W.~Li~\thanksref{addr10}  \and
        V.~Malyshev~\thanksref{addr12}  \and
        A.~Matsushita~\thanksref{addr10}  \and
        M.~Meucci~\thanksref{addr5}$^{ab}$ \and   
        S.~Mihara~\thanksref{addr13}  \and
        W.~Molzon~\thanksref{addr11} \and
        T.~Mori~\thanksref{addr10}  \and
        D.~Nicol\`o~\thanksref{addr1}$^{ab}$ \and
        H.~Nishiguchi~\thanksref{addr13}  \and
        A.~Ochi~\thanksref{addr14,e5}  \and
        W.~Ootani~\thanksref{addr10}  \and
        A.~Oya~\thanksref{addr10} \and
        D.~Palo~\thanksref{addr11} \and
        M.~Panareo~\thanksref{addr6}$^{ab}$ \and
        A.~Papa~\thanksref{addr1}$^{ab}$\thanksref{addr2} \and
        V.~Pettinacci~\thanksref{addr5}$^{a}$ \and
        A.~Popov~\thanksref{addr7,addr9} \and
        F.~Renga~\thanksref{addr5}$^{a}$ \and
        S.~Ritt~\thanksref{addr2} \and
        M.~Rossella~\thanksref{addr4}$^{a}$ \and
        A.~Rozhdestvensky~\thanksref{addr12}  \and
        S. Scarpellini~\thanksref{addr5}$^{ab}$ \and
        P.~Schwendimann~\thanksref{addr2} \and
        G.~Signorelli~\thanksref{addr1}$^{a}$ \and
        M.~Takahashi~\thanksref{addr14}  \and
        Y.~Uchiyama~\thanksref{addr13} \and
        A.~Venturini~\thanksref{addr1}$^{ab}$ \and        
        B.~Vitali~\thanksref{addr1}$^{a,}$\thanksref{addr5}$^{b}$ \and
        C.~Voena~\thanksref{addr5}$^{ab}$ \and   
        K.~Yamamoto~\thanksref{addr10}  \and
       R.~Yokota~\thanksref{addr10}  \and
        T.~Yonemoto~\thanksref{addr10}  
}

\institute{\JINR   \label{addr12}
           \and
             \INFNPi \label{addr1}
           \and
             \ICEPP \label{addr10}
           \and
             \INFNGe \label{addr3}
           \and
             \INFNPv \label{addr4}
           \and
             \INFNRm \label{addr5}
           \and
             \INFNLe \label{addr6} 
           \and
             \PSI \label{addr2}
            \and
             \ETHZ    \label{addr16}
            \and
             \INFNNa \label{addr17}
           \and
             \BINP   \label{addr7}
           \and
             \NOVST  \label{addr8}
           \and
             \NOVS   \label{addr9}
            \and
             \Liverpool  \label{addr15}
           \and
             \UCI    \label{addr11}
           \and
             \KEK    \label{addr13}
           \and
             \Kobe    \label{addr14}
}

\thankstext[*]{e1}{Corresponding authors:gianluca.cavoto@roma1.infn.it} 

\thankstext[$\dagger $]{e5}{Deceased} 

\date{Received: date / Accepted: date}

\maketitle

\begin{abstract}
The observation of a resonance structure in the opening angle of the electron-positron pairs in the  $^{7}$Li(p,\ee) $^{8}$Be reaction was claimed and interpreted as the production and subsequent decay of a hypothetical particle (X17). Similar excesses,  consistent with this particle, were later observed in processes involving $^{4}$He and $^{12}$C nuclei with the same experimental technique.
The MEG II apparatus at PSI, designed to search for the $\mu^+ \rightarrow \mathrm{e}^+ \gamma$ decay, can be exploited to investigate the existence of this particle and study its nature.
Protons from a Cockroft-Walton accelerator, with an energy up to 1.1 MeV, were delivered on a dedicated Li-based target. The $\gamma$ and the e$^{+}$e$^{-}$ pair emerging from the $^8\mathrm{Be}^*$ transitions were studied with calorimeters and a spectrometer, featuring a broader angular acceptance than previous experiments. 
We present in this paper the analysis of a four-week data-taking in 2023 with a beam energy of  1080 keV, resulting in the excitation of two different resonances with Q-value \SI{17.6}{\mega\electronvolt} and \SI{18.1}{\mega\electronvolt}.
No significant signal was found, and limits at \SI{90}{\percent} C.L. on the branching ratios (relative to the $\gamma$ emission) of the two resonances to X17 were set, $R_{17.6} <\SI{1.8e-6}{}$ and $R_{18.1} < \SI{1.2e-5}{}$.

\end{abstract}

\section{Introduction}
\label{sec:intro}

The study of the \ee~ opening angle  distribution in the $^7$Li(p,\ee)$^8$Be nuclear reaction revealed a significant anomaly with respect to the  expectations in an experiment conducted at the ATOMKI laboratory in Debrecen (Hungary) \cite{kr16}. 

 The \ee~pair is produced in the nuclear transition of the $^8$Be 17.6 MeV or the 18.1 MeV $J^{\pi}$ = 1$^+$ excited states\footnote{Indicated as $^8$Be$^{*}$(17.6) and $^8$Be$^{*}$(18.1) hereinafter.} to a lower-lying excited state or to the ground state (see Fig.~3 in \cite{TILLEY2004155}). Such a process is commonly known as Internal Pair Conversion (IPC) and takes place with a $10^{-3}$ probability with respect to the $\gamma$-ray  emission process  $^7$Li(p,$\gamma$)$^8$Be. 
 According to an accepted nuclear physics models
(\cite{SCHLUTER1981327}\cite{PhysRev.76.678}) the \ee~opening angle  \thetaee~has a distribution that is monotonically  decreasing from its  maximum close to 10$^\circ$ to a null value at 180$^\circ$. 
 A departure from this falling distribution might imply the presence of a new physics degree of freedom such as the creation of an intermediate new particle.  The experimental evidence consists of an excess of events with an invariant mass around 17 MeV when exciting the $^8$Be$^{*}$(18.1) state. \sloppy It was further confirmed by other refined measurements at ATOMKI and additional studies of the $^3$H(p,\ee)$^4$He   ~\cite{kr141,kr17,Krasznahorkay:2019lyl} and  $^{11}$B(p,\ee)$^{12}$C  ~\cite{AtomkiCarbon} reactions. A  hypothetical particle, consistent with their  results  - now commonly referred to as X17 - was introduced in several theoretical attempts to explain the anomaly.  Standard Model or nuclear physics effects unaccounted for previously have also been suggested. 
 For a review of the theoretical and experimental status of the study of this anomaly, see~\cite{Alves:2023ree}.
 This situation clearly urges a diversified experimental program to further test its existence and clarify its nature. 
A recent experiment at VNU, Vietnam~\cite{Anh2024}, utilizing apparatus and analysis methods very similar to those used by ATOMKI, confirmed the anomaly. Additionally, an experiment at JINR~\cite{Abraamyan2023} reported an excess in the same mass range, albeit through a different process.

The MEG II collaboration detector \cite{MEGII:2023fog} employs the $^7$Li(p,$\gamma$)$^8$Be reaction for its LXe calorimeter calibration  \cite{MEG:2011rgj} and is in a position to reproduce ATOMKI's results with a different detection technique. 
In MEG II, the nuclear reactions can be  excited with a proton beam impinging on a thin Li-based target with a beam energy variable between a few hundred keV up to more than 1000 keV. The two main resonances at 440 keV and 1030 keV can then be accessed to form  
$^8$Be$^{*}$(17.6) and $^8$Be$^{*}$(18.1) states respectively.
While the  ATOMKI apparatus is confined to a plane orthogonal to the proton direction and uses  scintillators and a vertex detector to measure the direction and momentum of the charged particles without a magnetic field,   MEG II employs a magnetic spectrometer with a cylindrical drift chamber offering larger angular coverage \cite{DCHMEGperformances}\cite{Baldini_2018}.

 In this paper, we report the results of the search for the hypothetical X17 particle  performed on data taken in February 2023 during a dedicated data-acquisition campaign. The energy of the protons on target was such as to explore simultaneously both resonances and therefore to investigate the production of the X17 particle  from both $^8$Be$^*(17.6)$ and $^8$Be$^*(18.1)$ transitions to ground state.
 In Sec.~\ref{sec:app}, the MEG II apparatus is shortly described with focus on the subsystems relevant to the measurement and the data-taking strategy is outlined. In Sec.~\ref{sec:anal}, the procedure to reconstruct the \ee~pairs and the data selection are outlined. A maximum likelihood technique is employed to extract the X17 yield  and discriminate it from the abundant IPC and $\gamma$ pair-conversion background events. In Sec.~\ref{sec:branch} the branching fraction  results with respect to the $\gamma$-ray production are reported.

\section{The MEG II apparatus for the X17 search and data sample}
\label{sec:app}
\subsection{The MEG II apparatus}  
\label{subsec:app}

\begin{figure}[htbp]
    \centering
    \includegraphics[width=0.40\textwidth]{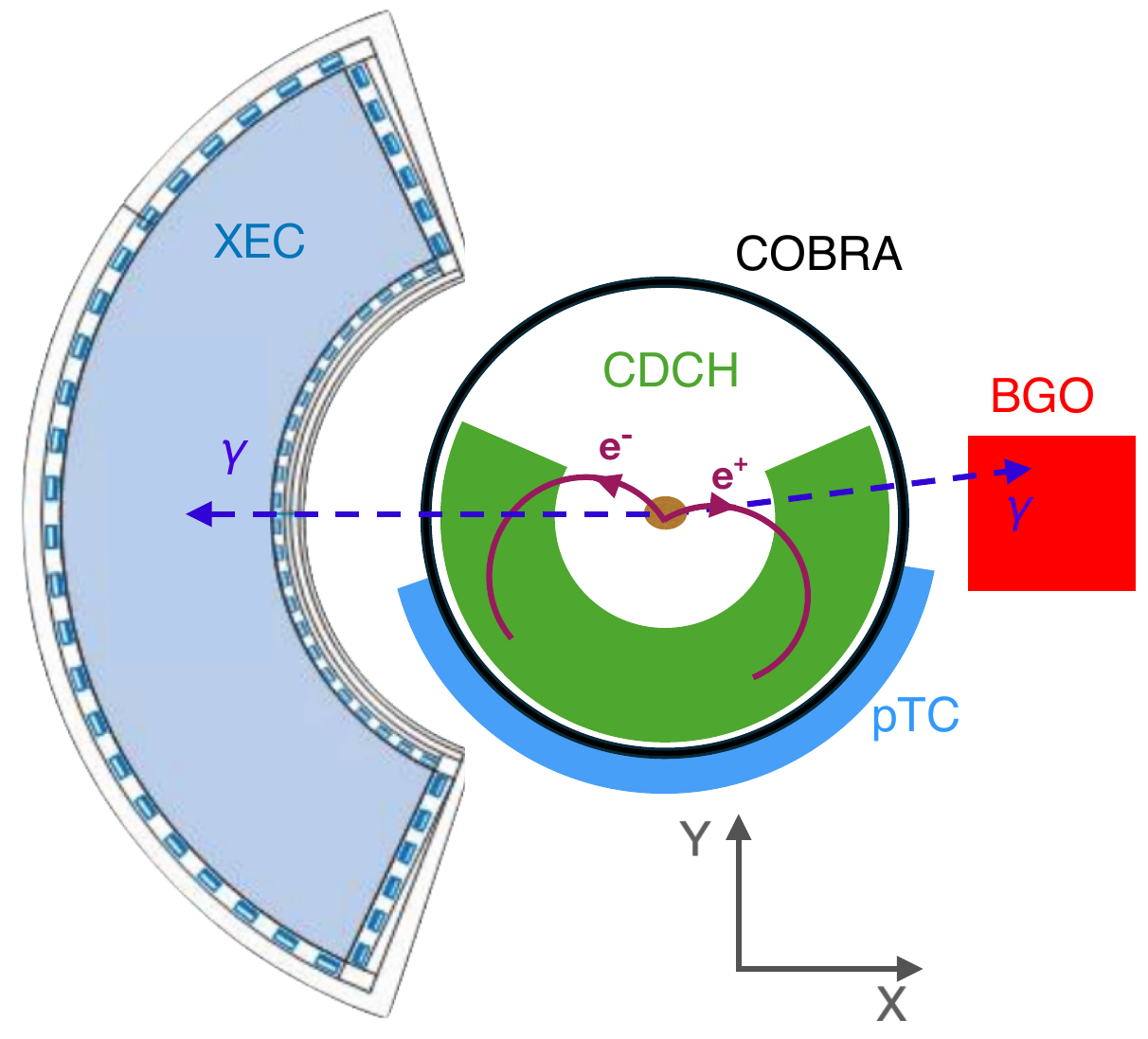}
        \includegraphics[width=0.48\textwidth]{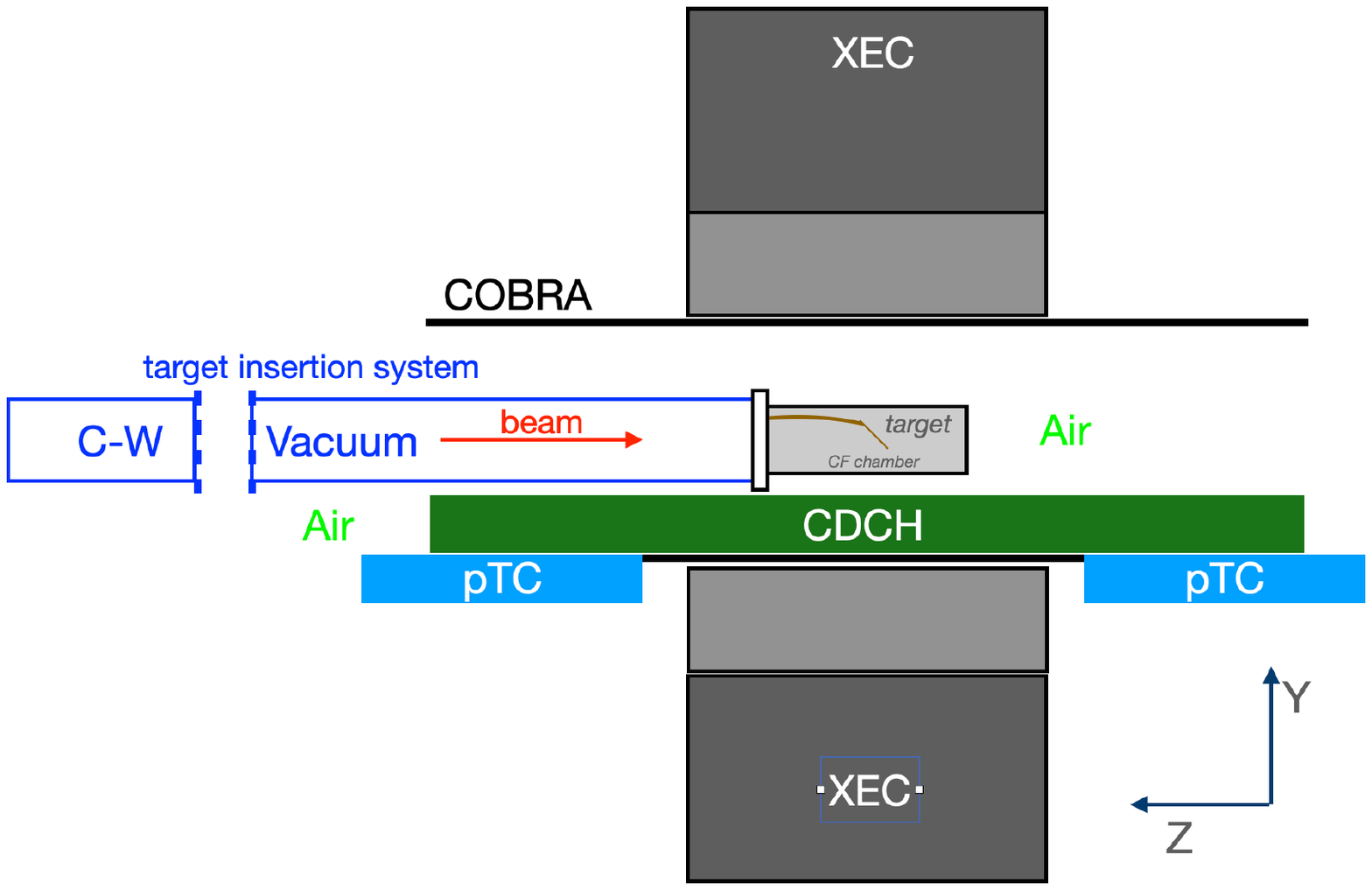}
    \caption{Representation of the MEG II apparatus employed for this work.  The  Cockroft-Walton proton accelerator is located  at a $z$ position several meters away from the COBRA center. 
    }
    \label{fig:apparatuspic1}
\end{figure}

 The MEG II experimental apparatus was employed for this search. A full description can be found elsewhere \cite{ref:MEGIIdesign,MEGII:2023fog}. It was designed and optimized for the search of the ultra-rare decay $\mu^+ \rightarrow \mathrm{e}^+ \gamma$, nevertheless, some of its subsystems are highly suitable for the X17 search.  They include a spectrometer consisting of a multi-wire cylindrical drift chamber (CDCH) made up of 1728 separate cells and two 256 scintillator tiles' arrays readout by SiPMs to measure the charged particles timing - one located upstream of the target and the other downstream (pTC) - all housed within a superconducting solenoid aligned with the beam axis and featuring  a gradient magnetic field (COBRA) of 1.27 T at its center (Fig.~\ref{fig:apparatuspic1})  
  
 A right-handed, Cartesian coordinate system is adopted, with the $z$-axis along the direction opposite to the proton beam and the y-axis vertical and pointing upward.
 The origin of this coordinate system is located at the center of the COBRA magnet (Fig.~\ref{fig:apparatuspic1}). Moreover,  a polar spherical coordinate system with $\theta$ defined as usual relative to the beam (z) axis is also employed.

  A 1 MV Cockroft-Walton accelerator (CW) from High Voltage Engineering Europa is used to produce both proton ($\mathrm{H}^+$) and  $\mathrm{H}_2^+$ ions with maximal kinetic  energy ($E_b$) slightly above 1 MeV and with current ranging from 1 $\mu$A to several tens of $\mu$A  \cite{ADAM201119}. In the MEG II experiment, it  is routinely used in the regular  $\mu^+ \rightarrow \mathrm{e}^+ \gamma$ data taking to calibrate a LXe detector (XEC) with $\gamma$-rays from the same $^7$Li(p,$\gamma$)$^8$Be nuclear reaction, exciting the 440 keV resonance.
   The beam used in the X17 search DAQ period was, in fact, a mixture of 75$\%$ $\mathrm{H}^+$ and 25$\%$ $\mathrm{H}_2^{+}$ ions, accelerated to the same kinetic energy. The composition of the beam mixture  was measured with a Faraday cup in a dedicated test.
  
  The CW vacuum beam pipe is connected to the center of the COBRA magnetic field where a LiPON (lithium phosphorus oxynitride) target with 7~$\mu$m average thickness is installed at the center of COBRA. LiPON was chosen for its chemical stability and was deposited by sputtering on a 25 $\mu$m-thick copper substrate. It is mechanically connected to a copper holder designed to ease heat dissipation through the vacuum pipe flanges.  The quality of the LiPON deposit was examined with a scanning electron microscope. It revealed a rough surface which can impact the proton energy distribution and some negligible surface contamination. 
  
  To minimize the  \ee~multiple scattering and the $\gamma$-ray conversions,  the target region is encapsulated in a  400~$\mu$m-thick carbon fibre cylinder. 
  However, with  the current setup, \ee~particles cross an air layer with thickness varying between 12 and 16 cm  before reaching the CDCH volume. 

\begin{figure}[htbp]
    \centering
    \includegraphics[width=0.48\textwidth]{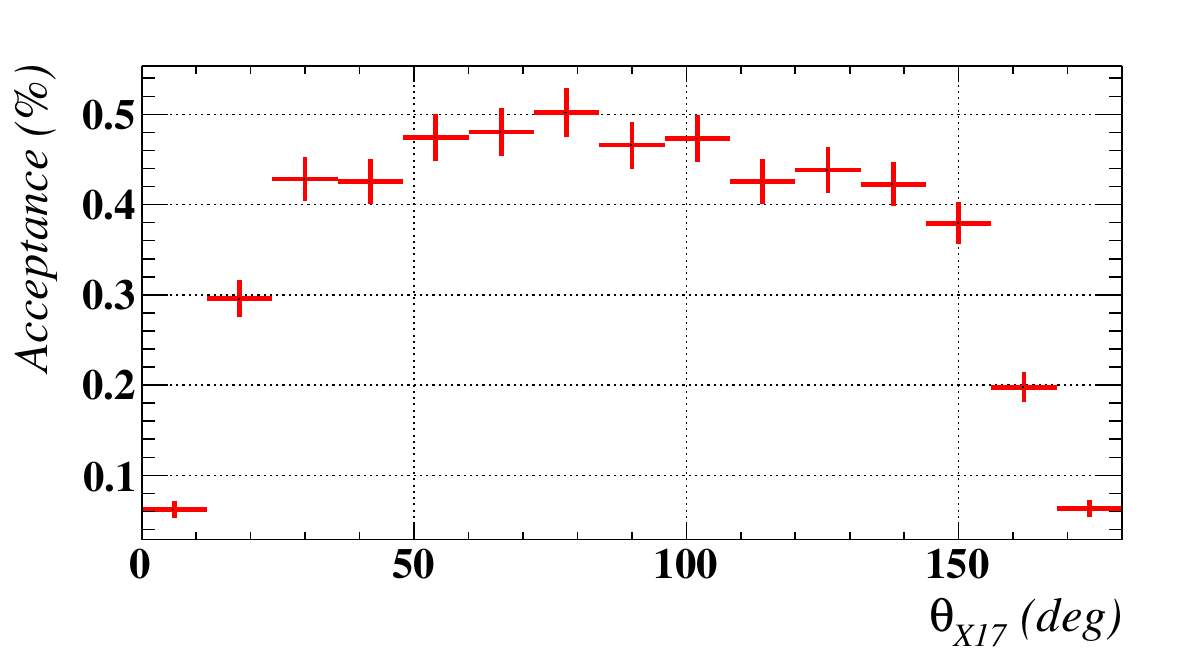}
    \caption{Acceptance for X17 signal events as a function of the polar angle of the X17 momentum, estimated with a Monte Carlo simulation of the apparatus.}
    \label{fig:acceptance}
\end{figure}

  After hitting the CDCH cells, the  e$^+$ and  e$^-$  particles are eventually detected by one of the two pTC sectors. A trigger to select signal X17 events is based on the coincidence of hit scintillators in the pTC and hit cells in the CDCH. 
  During the X17 data taking a 4x4 Bismuth Germanate (BGO) crystal matrix, $4.6 \times 4.6 \times 20 \; \rm{cm^3}$ each,  was used to detect $\gamma$-rays from the $^7$Li(p,$\gamma$)$^8$Be process. It put forward good stability of both the target and the beam. Fig.~\ref{fig:acceptance} shows  the acceptance of X17 MC-computed signal events as a function of the X17 direction with respect to the beam axis. It reveals how MEG II might be in a position to explore X17's quantum numbers~\cite{Viviani:2021stx}.

\subsection{The data sample} 
\label{subsec:data}

The COBRA magnetic field was set to 15$\%$ of the nominal value used during the $\mu^+ \rightarrow \mathrm{e}^+ \gamma $ data-taking, in order to accept the lower momenta of e$^+$ and  e$^{-}$   in the spectrometer. 
The BGO  was used for background studies, beam and target stability monitoring and normalization purposes. Moreover, a 3-inch Lanthanum Bromide crystal and the XEC were employed in some auxiliary measurements. 
The CW was operated at a kinetic energy $E_b = 1080~\mathrm{keV}$ with currents ranging from 8 $\mu$A to 11 $\mu$A. 

 As previously stated, the X17 search trigger is a coincidence between pTC and CDCH conditions:
 \begin{itemize}
      \item pTC trigger fired when at least one tile is hit
     \item CDCH trigger fired when a number of hits are simultaneously reconstructed on both wires' readout sides. The actual number was optimized based on signal-to-background efficiency from MC simulations.

 \end{itemize}
 
  The trigger rate was about 35 Hz for a proton current close to 10~$\mu$A.
The MEG II  WaveDAQ \cite{FRANCESCONI2023167542} system was employed for the trigger implementation and data readout based on Domino Ring Sampler 4 chips for the analog sampling of the detector signals at frequency larger than 1 GHz.

A stable $\gamma$-ray rate and no significant target deterioration were observed using BGO data. A Gaussian-shaped beam spot with a transverse $\sigma$  of 3
mm~ was tuned using a proton-induced fluorescent quartz target, which was imaged with a CCD camera before data-taking.
By varying three  dipolar fields along the beamline, the beam spot was centred to cover the target area.
When  the H$_2^{+}$ ions  enter the target, their two protons interact with Li nuclei each with roughly half of the H$_2^{+}$ kinetic energy. 

\begin{figure}
\centering
\resizebox{0.99\columnwidth}{!}{%
 \includegraphics{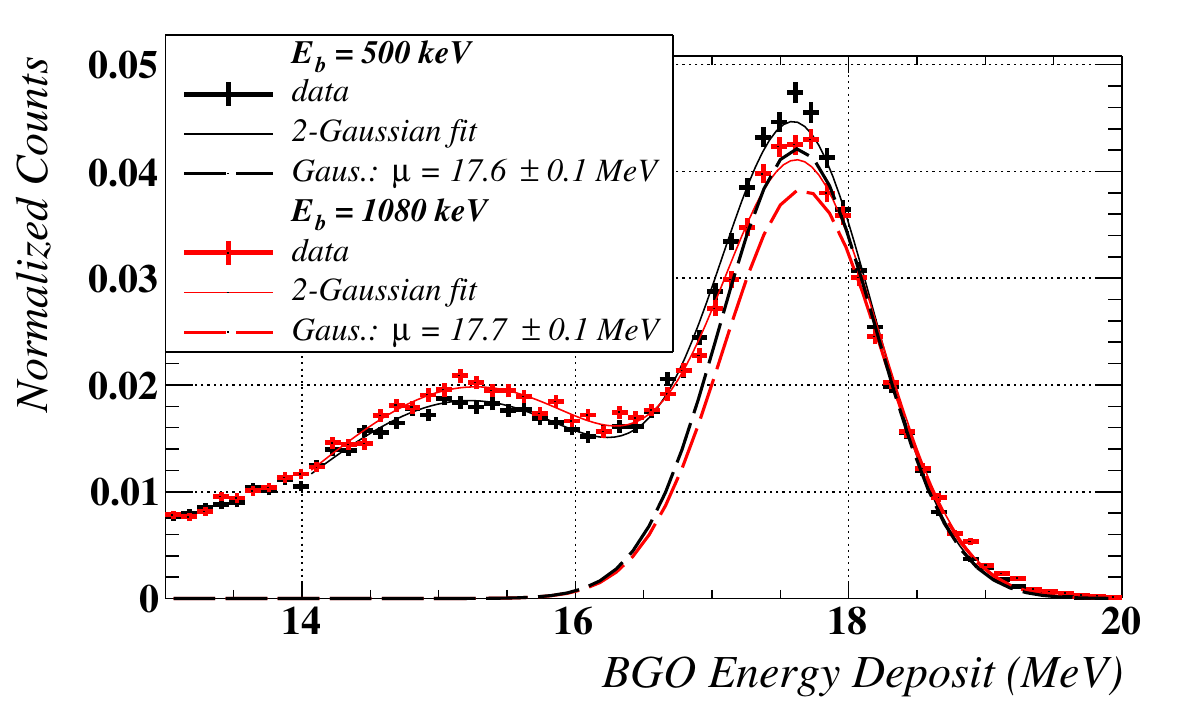} }
\caption{BGO detector energy deposit distribution from two datasets with $E_b = 1080$~keV (red) and $E_b = 500$~keV (black). The two-Gaussian fit and the 18.1~MeV component for each distribution are displayed as a full and a dashed line respectively. A 4\% energy resolution of the detector dominates the higher energy line width.}
\label{fig:BGOevidence}       
\end{figure}

 The H$_2^+$ presence   in the  CW beam 
 causes both 440~keV and 1030~keV Be resonances to be populated, leading to the excitations of both 17.6~MeV   and 18.1~MeV Be$^{*}$ states. An estimate of the relative proportions of the two resonances is necessary for the X17 search. 
Figure~\ref{fig:BGOevidence} shows the BGO energy deposit from two datasets, one with $E_b$ = 500~keV and the other with $E_b$ =  1080~keV. At $E_b$ = 500~keV, one can see the 17.6~MeV and 14.6~MeV $\gamma$-ray lines associated with the Be$^*$(17.6) transition to the ground state and to the first excited state, respectively. In the case of a pure $\mathrm{H}^+$  beam, the photon line is expected  to be approximately 400~keV higher at $E_b$ =  1080~keV than at $E_b$ =  500~keV.  
However, the observed relative shift of the two distributions (with $E_b$ =  1080~keV and $E_b$ =  500~keV) is found to be 100~keV: the H$_2^+$ species leads to a dominant contribution of the  440~keV  resonance in the $E_b = 1080$~keV dataset. Moreover, protons in the target lose energy (about \SI{40}{\kilo\electronvolt/\micro\meter} at \SI{1000}{\kilo\electronvolt} kinetic energy), resulting in the presence of events with intermediate energy (averaging around 17.9 MeV), which must be included in the analysis.

Consequently, the maximum likelihood technique used to extract the signal (Sec.~\ref{subsec:fitprocedure}) requires modelling events originating from 440~keV and 1030~keV resonances and from intermediate energies. Events from the 1030 keV resonance are estimated to account for approximately 20\% of the total, and their fraction is extracted using the maximum likelihood fit itself.


A sample of 75M events at $E_b = 1080$~keV was collected during a dedicated four-week data taking campaign in February 2023.




\section{Analysis Method}
\label{sec:anal}

\subsection{Signal model and Monte Carlo simulation} 
\label{subsec:MC}


A detailed Monte Carlo (MC) simulation based on the Geant4 toolkit \cite{Agostinelli2003250} within the MEG II software framework  is employed to generate $\gamma$-rays  or IPC \ee~pair interacting with the material of the experimental apparatus (target, beam pipe, detectors). In each simulated event, the $\gamma$-ray or the IPC \ee~pair   originating from  the nuclear process is propagated through  the MEG II detector.

 The Zhang-Miller (Z-M) model   \cite{Zhang:2017zap} accounts for photon anisotropy and multipole interferences. The $\gamma$-ray and IPC \ee~generations follow the Z-M cross-section. The real and virtual $\gamma$-rays  originating from the de-excitation of the $^8$Be nucleus to ground state were produced over the full solid angle with an energy given by $E_p$ + $E_{th}$ where $E_{th}$ = 17.25 MeV is the sum of the masses of $^7$Li and of the proton minus the  $^8$Be ground state mass~\cite{TILLEY2004155}. $E_p$ was generated  ranging from 300 keV to 1080 keV. A similar set of events was produced for the   de-excitation from $^8$Be$^*$  to  the 3.0 MeV first  excited state. Though the Z-M  model   originally only considered the transition to the ground state in their work, we extended it to the transition to the first excited state. 

The  $\gamma$-rays  all have sizeable probability to produce charged particles in the apparatus material via Compton scattering or pair production. These $\gamma$-ray events are generically classified as External Pair Creation (EPC)  events and a MC sample is generated accordingly.

While the number of generated IPC MC events is outnumbering the IPC events in data, the EPC MC sample was a factor 20 smaller than the number observed in data, being the EPC background less relevant in the signal region.


The X17 signal (with masses between 16.5 and 17.1~MeV/c$^2$) is generated isotropically from the de-excitation of both $^8$Be$^*$(17.6) and $^8$Be$^*$(18.1). The $^8$Be nucleus is assumed at rest in the laboratory frame, thus neglecting its small boost. 
The X17 particle is assumed to instantaneously decay in an \ee~pair at a random point within the Gaussian-shaped beamspot.

 \subsection{Reconstruction}  
\label{subsec:reco}


Electron and positron  tracks in the CDCH and pTC  are reconstructed with the MEG II track finder and then fitted with a Kalman filter algorithm, using the standard MEG II magnetic field map scaled by a factor of 0.15 and the detector material simulation. 

\begin{figure}[htbp]
    \centering
    \includegraphics[width=0.48\textwidth]{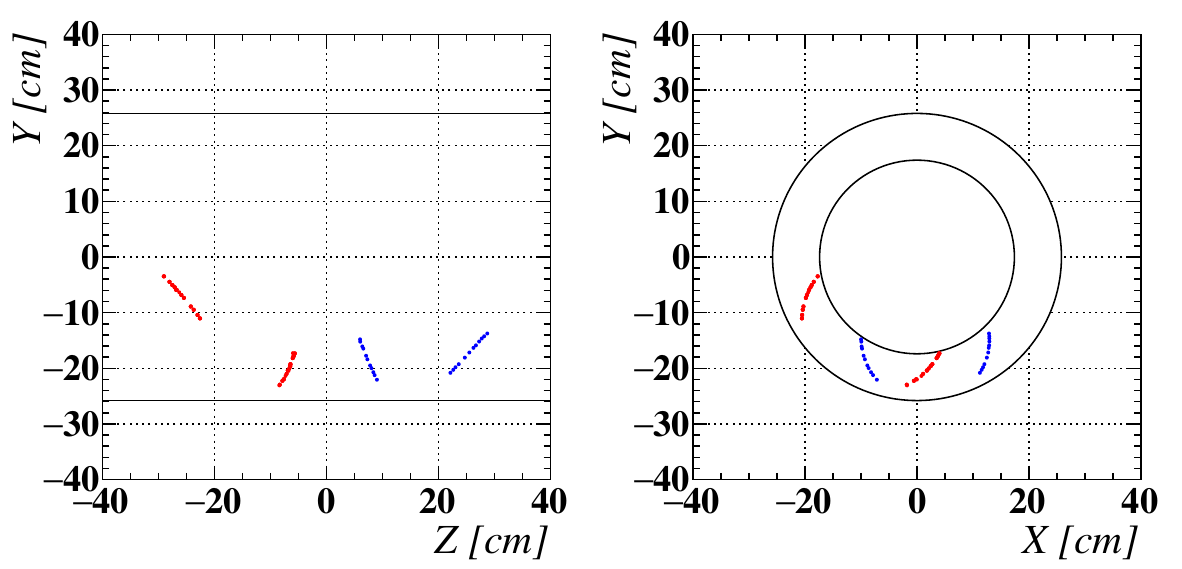}
    \caption{CDCH hits from a signal-like data event associated to reconstructed tracks. Positron and electron  hits fitted to a track in red and blue respectively.}
    \label{fig:goodevent}
\end{figure}

All tracks are propagated back to the point of closest approach (POCA) to the z-axis. The z-coordinate of the POCA is referred to as $z_{vtx}$.  Lists of reconstructed positron and electron tracks are given as input to a stringent track and pair selection described in Sec.~\ref{subsec:eventsel}. In each event, only the \ee~pair of tracks with the lowest momentum uncertainty, as estimated from the track fit,  is  used in the analysis. The momentum of each track is then extrapolated to the z-axis POCA. 
Figure~\ref{fig:goodevent} shows in the (Z,Y) (left) and (X,Y) (right) planes the fitted hits of a signal-like data event where a positron track (red hits) and an electron track (blue hits) are found.


   \subsection{Event selection} 
\label{subsec:eventsel}
    

 In some cases,  the reconstruction algorithm can yield fake tracks: hits from an electron particle are erroneously fitted to a positron track and vice versa. Pair of tracks including a fake track, referred to as fake pairs, have large angular openings, close to 180$^\circ$. To preserve a good signal sensitivity, a balance was found between fake track rejection and good track efficiency. Fake tracks were simulated and in general found to be short, to have small hit density, their hits close to $z=0$, and to be emitted orthogonally to the beam axis. 
  
 A detailed  selection  on both single tracks  and on the pair of tracks was developed and is summarized in Tab.~\ref{tab:track-selection}. The variables used in the selection include  $n_{hits}$ representing the number of  CDCH hits used in a track fit; $z_{b}$, $z_{f}$ and $z_{l}$ being  the $z$-coordinate of the beamspot center, of the first hit and of the  last hit used in the track fit respectively. $T_{0f}$ and $T_{0l}$ represent the first and last hit timings according to the track fit and the propagation length is defined as  the distance travelled by the particle from the z-axis POCA to the first hit in the CDCH. We defined  the hit density $\mu_{hit}$ as the number of  hits per cm along the track trajectory and the  track score as $n_{hits} + 10~\mu_{hit}$. The average of the $z$-coordinates   of all the fitted hits assigned to  a track is referred to as $z_{mean}$. The consecutive hits distance std is defined as the standard deviation of the distribution of distances between all pairs of neighbouring fitted hits.

\begin{table}[htbp]
    \centering
    \caption{List of selection criteria on single and pair of tracks.}
    \begin{tabular}{|c|c|c|c|}
        \hline
        \textbf{Selection}  \\
        \hline
        $n_{hits} \geq 10$   \\
        \hline 
        $|z_{vtx}-z_{b}| \leq 2.5~\mathrm{cm}$  \\
        \hline 
        $T_{0l} - T_{0f} \geq 0$  \\
        \hline 
        $(z_{l} - z_{f}) \times sgn(z_{f})\geq 0$   \\
        \hline 
        propagation length $\geq 35~\mathrm{cm}$   \\
        \hline 
        \begin{tabular}{l} if $10\leq n_{hits} \leq16$,\\$\mu_{hit} \geq 1.1~\mathrm{hits/cm}$ \end{tabular}  \\
        \hline
        \begin{tabular}{l} \textit{if $\mu_{hit}> n_{hits}/12-2/3$}:\\$\mu_{hit} \geq 0.8~\mathrm{hits/cm}$ \\track score $\geq 20$ \end{tabular}  \\
        \hline

        Consecutive hits distance std $<0.9~\mathrm{cm}$  \\
        \hline        $\vert z_{f}\vert \geq 2.5~\mathrm{cm}$   \\
        \hline

        $z_{mean} \times (\theta-90^\circ) < 0$   \\
        \hline

 \begin{tabular}{l}No hits in common between e$^+$ and e$^-$ tracks\end{tabular}  \\
        \hline 

        \begin{tabular}{l} \ee vertices distance $<~3~\mathrm{cm}$\end{tabular}   \\

        \hline
    \end{tabular}
    \label{tab:track-selection}
\end{table}

\begin{figure}
\centering
\resizebox{0.99\columnwidth}{!}{%
 \includegraphics{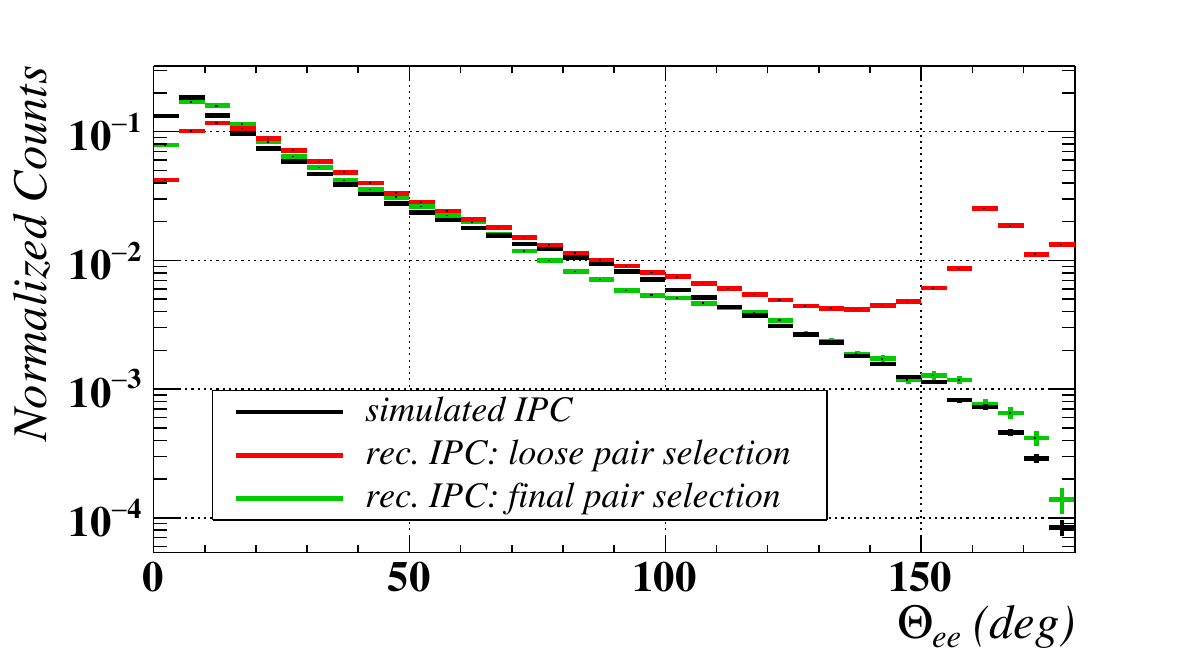} }
\caption{Distributions of \thetaee~from IPC MC events, as predicted by the Z-M model (black), and following detector simulation and reconstruction, before (red) and after (green) applying the selection described in Tab.~\ref{tab:track-selection}. A significant excess of fake pairs is observed at large angles prior to the final selection.}
\label{fig:clooseVStightIPC_log}       
\end{figure}

The effect of the track selection is observed in Fig.~\ref{fig:clooseVStightIPC_log}. The reconstructed IPC MC angular opening (\thetaee)~distribution is displayed under two conditions, after a loose track selection and after applying  the  more stringent selection from Tab.~\ref{tab:track-selection}. The fake pairs contribute to a significant structure above 150$^\circ$ which could reduce the signal sensitivity. The final selection recovers the IPC MC monotonous shape though some discrepancies related to apparatus acceptance are observed. 

 In Tab.~\ref{tab:numbers}  the MC-estimates of the  trigger and  reconstruction efficiency  for a 16.9~MeV/c$^2$ X17 signal from $^8$Be$^*$(17.6) decay, IPC and EPC backgrounds are reported. The quoted numbers are obtained from MC simulations and corrected based on data to account for some observed inefficiency. The trigger efficiency is given with respect to the \ee~production for X17 and IPC and with respect to the $\gamma$-ray production for EPC.


\begin{table*}[h]
    \centering
    \begin{tabular}{|c|c|c|c|c|c|c|}
         \hline
    &  X17  & IPC    & EPC          \\
    &  $16.9~\mathrm{MeV/c}^2$& 18 MeV   & 18 MeV          \\
     \hline
         trigger selection eff.  & 16\% & 4.7\%&   0.026\%   \\      \hline
          e$^+$ selection eff. (wrt trg)    & 24\%    & 26\%&   13\%  \\      \hline
               \ee~selection eff. (wrt trg)          & 2.5\% & 2.3\%&   0.6\%  \\      \hline
      \thetaee~resolution [deg] & $5.6\pm0.2$ & $5.5\pm0.1$& // \\ \hline
      $E_{sum}$ resolution [MeV] & $0.58\pm0.02$ & $0.69\pm0.01$& // \\ \hline

        \end{tabular}
    \caption{MC-based trigger, positron selection efficiency, pair selection efficiency and resolutions of the  analysis variables for  the X17 signal with  $m_0$ = 16.9~MeV/c$^2$ from Be$^{*}(17.6)$ de-excitation, the IPC and EPC  backgrounds.}
    \label{tab:numbers}
\end{table*}

\subsection{Analysis variables} 
\label{subsec:analvar}


Once the best \ee~pair is selected, the momenta of each track are computed at z-axis POCA. Two analysis variables are then extracted:
\begin{itemize}
    \item  \thetaee $= \mathrm{cos}^{-1} \left(\frac{\textbf{p}_+\cdot \textbf{p}_-}{|\textbf{p}_+||\textbf{p}_-|}\right) $, the angular opening between the e$^+$ and e$^-$ momenta 
    \item $E_{sum} = E_+ + E_-$, the sum of the e$^+$ and e$^-$ energies
\end{itemize}
Tab.~\ref{tab:numbers} gives the resolutions for these variables as estimated from MC simulations.

\begin{figure}
\centering
\resizebox{0.99\columnwidth}{!}{%
 \includegraphics{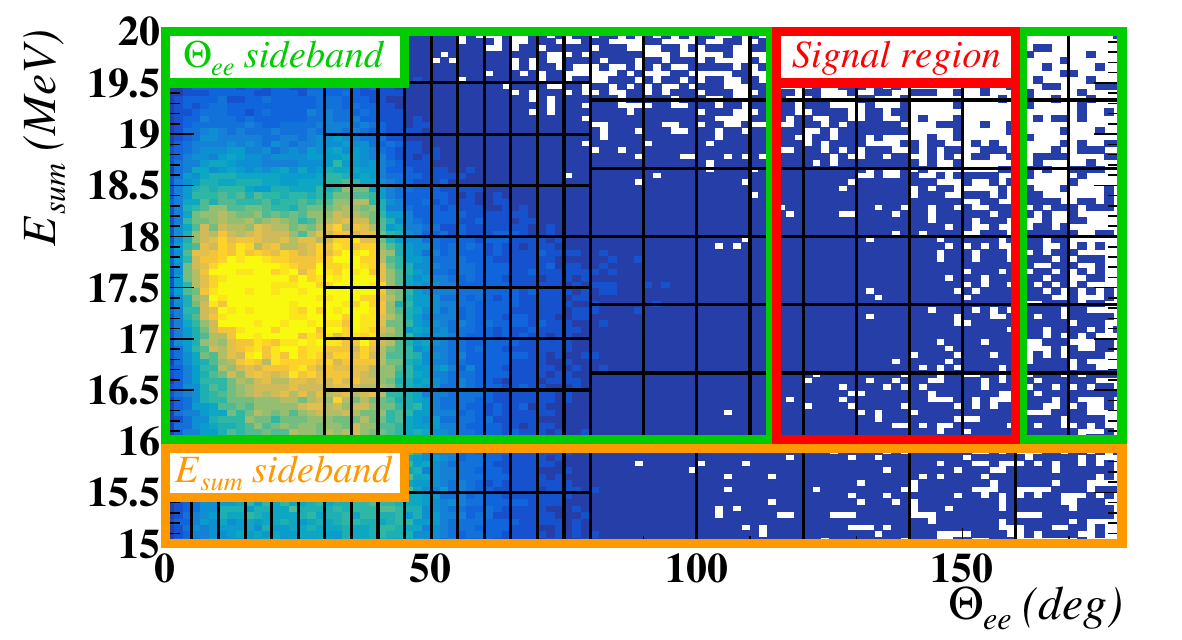} }
\caption{February 2023 dataset  $E_{sum}$ vs ~\thetaee~plane. The ~\thetaee~and E$_{sum}$ sidebands are shown in purple. The signal region after unblinding is shown in red here. The binning grid used in Sec.~\ref{subsec:fitprocedure} is overlaid in black.}
\label{fig:esumVSangle_sidebands}       
\end{figure}

The dataset is analyzed making use of these two sole variables. As illustrated in Fig.~\ref{fig:esumVSangle_sidebands}, a blinded box was defined in the ($E_{sum}$, $\Theta_{ee}$) 2D-plane for pairs with $16~\mathrm{MeV} \leq \mathrm{E}_{sum} \leq 20~\mathrm{MeV}$ and $115^\circ \leq \Theta_{ee} \leq 160^\circ$. This region was chosen based on Atomki's X17 mass estimate and blinds any excess arising from Be$^{*}(17.6)$ or Be$^{*}(18.1)$ transitions. Two sidebands were also defined in this plane. The \thetaee~sideband gives access to the energy sum signal region while the  E$_{sum} $ sideband permits studying the full \thetaee~range. 

Based on studies performed in the sidebands, events in the region  $\SI{16}{\mega\electronvolt} < E_{sum} < \SI{20}{\mega\electronvolt}$, $\Theta_{ee} < \SI{30}{\degree}$ are particularly sensitive to  the proton beam position on the target and to the fraction of $\mathrm{H_2^+}$ in the beam, and are consequently excluded from the analysis. For the same reason, events in the region $\Theta_{ee} < \SI{50}{\degree}$, $\theta_{X17} < \SI{80}{\degree}$ are also removed. Both regions are far from the signal one and hence this selection does not have any impact on the signal acceptance.

\begin{figure*}
\centering
 \resizebox{1.99\columnwidth}{!}{%
 \includegraphics{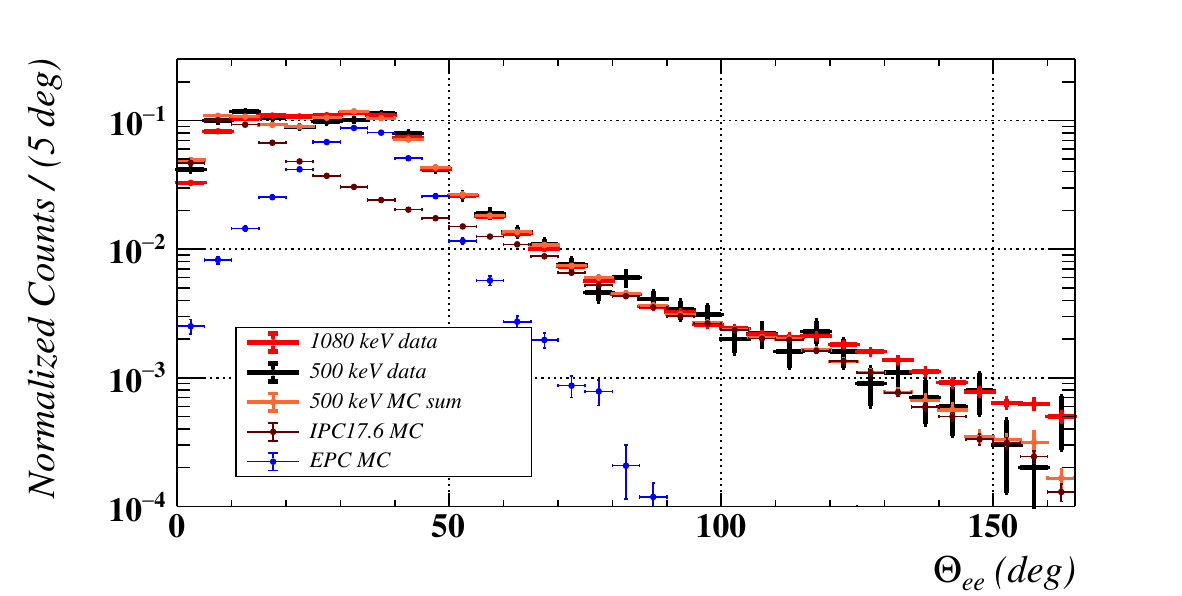} }
\caption{\thetaee~distribution from $E_b = 1080$~keV 2023 dataset (red) and from a small $E_b = 500$~keV dataset (black) for $16~\mathrm{MeV} \leq \mathrm{E}_{sum} \leq 20~\mathrm{MeV}$. The 500~keV data are fitted with a 500~keV MC sum (orange) composed of both EPC MC (blue) and IPC (brown) from the Be$^{*}(17.6)$ de-excitation.}
\label{fig:angularevidence}       
\end{figure*}

In order to evaluate the validity of the Z-M IPC model, the \thetaee~distribution of a small dataset taken at $E_b = 500$~keV is fitted with a 500~keV sum of EPC   and IPC   MC following Z-M model relative to the $^8$Be$^{*}$(17.6) de-excitation.   They are  compared  in Fig.~\ref{fig:angularevidence} and show good agreement. Though statistically limited, this dataset can be well explained by the sole EPC and IPC backgrounds, though EPC is negligible in the signal region.

 This  dataset is also compared to the February 2023 $E_b = 1080$~keV data. Though very close in shape up to \thetaee~$= 100^\circ$, the latter dataset then diverges. The flatter shape of the E1-enriched IPC from the Be$^{*}(18.1)$   state with respect to the M1-dominated IPC from Be$^{*}(17.6)$  explains the systematic discrepancy\footnote{For a definition of multipole M1 and E1  see \cite{Blatt:1952ije}}
 This provides further evidence for the minority presence of events from the 1030~keV resonance in  the February 2023 dataset. The exact proportion of  Be$^{*}(18.1)$ is estimated through the maximum likelihood fit in Sec.~\ref{subsec:fitprocedure} making use of the \thetaee~shape.
 

\subsection{Maximum likelihood fit }
\label{subsec:fitprocedure}

The X17 search is carried out based on a binned likelihood function. The template histograms modeling both the backgrounds and the signal are generated from the high-fidelity Monte Carlo of the MEG II detector and validated in the sidebands prior to unblinding.

Both resonances at \SI{440}{\kilo\electronvolt} and \SI{1030}{\kilo\electronvolt} are expected to be excited. In addition, due to the energy loss of the  proton   in the target, a contribution to IPC from the intermediate non-resonant region could be expected.

A different background model is applied for IPC events produced in three distinct proton energy bins (2 resonant, \SI{18.1}{\mega\electronvolt} and \SI{17.6}{\mega\electronvolt} + 1 non-resonant, \SI{17.9}{\mega\electronvolt}), resulting in six separate populations with energies  \SI{14.6}{\mega\electronvolt}, \SI{14.9}{\mega\electronvolt}, \SI{15.1}{\mega\electronvolt}, \SI{17.6}{\mega\electronvolt}, \SI{17.9}{\mega\electronvolt}, \SI{18.1}{\mega\electronvolt},  including therefore the de-excitations occurring both to the $^8$Be ground state or to the \SI{3}{\mega\electronvolt} $^8$Be first excited state. Two signal hypotheses are included in the likelihood fit, one for each resonance.
Analogously, two EPC backgrounds are included in the fit to model the $\gamma$-ray creation from the two $\gamma$-raylines (de-excitation to the first excited and to the ground state). In this case no  difference can be resolved in the spectra of the EPC resulting from different proton energies according to our simulation.
In addition to the physical backgrounds, a template for fake pairs is included, accounting for fake \ee~pairs which survive the event selection.
The systematic effects introduced by the limited statistics of the Monte Carlo samples is accounted for by the simplified Beeston-Barlow approach proposed in   \cite{Dembinski2022}.

The X17 signal search is carried out at the two resonances. The number of signal events from each resonance is given by the branching ratio\footnote{Hereinafter, all branching ratios are intended for beryllium decays to ground state.} $\mathcal{B}(^8\mathrm{Be}^*(Q) \to ^8\mathrm{Be} + \mathrm{X17})$, with $Q$ either \SI{17.6}{\mega\electronvolt} or \SI{18.1}{\mega\electronvolt}, and the number of the corresponding IPC population, $\mathcal{N}_{\mathrm{IPC}Q} $:
\begin{eqnarray}
    \mathcal{N}_{\mathrm{X17}Q} &=& \mathcal{N}_{\mathrm{IPC}Q} \times R_{\mathrm{IPC}Q} \times k(m_{\mathrm{X17}})
\end{eqnarray}
where $R_{\mathrm{IPC}Q}$ is the ratio between  the branching ratio to the signal and the branching ratio to the IPC for the excited states at $Q$: 
\begin{eqnarray}
    R_{\mathrm{IPC}Q} &=& \frac{\mathcal{B}(^8\mathrm{Be}^*(Q) \to {^8\mathrm{Be}} + \mathrm{X17})}{\mathcal{B}(^8\mathrm{Be}^*(Q) \to {^8\mathrm{Be}} + \mathrm{e^+e^-})}
\end{eqnarray}
and:
\begin{eqnarray}
    k(m_{\mathrm{X17}}) = \frac{\epsilon_{\mathrm{X17}}(m_{\mathrm{X17}})}{\epsilon_{\mathrm{IPC}}}
\end{eqnarray}
is the ratio between  the efficiency of the apparatus to the signal and the efficiency to the IPC. This efficiency includes the relative geometrical acceptance on which an uncertainty of \SI{20}{\percent} is estimated by comparing data and Monte Carlo.

The likelihood function is explicitly parameterised in terms of the branching ratios relative to the $\gamma$-ray emission,
\begin{equation}
    R_Q = \frac{\mathcal{B}(^8\mathrm{Be}^*(Q) \to {^8\mathrm{Be}} + \mathrm{X17})}{\mathcal{B}(^8\mathrm{Be}^*(Q) \to {^8\mathrm{Be}} + \gamma)}\; .    
\end{equation}
They are calculated as $R_Q = R_{\mathrm{IPC}Q} \times \Gamma_{Q}$, with:
\begin{equation}
    \Gamma_{Q} = \frac{\mathcal{B}(^8\mathrm{Be}^*(Q) \to {^8\mathrm{Be}} + \mathrm{e^+e^-})}{\mathcal{B}(^8\mathrm{Be}^*(Q) \to {^8\mathrm{Be}} + \gamma)}\; .
\end{equation}

To facilitate comparison with the results from ATOMKI but enable potential rescalings based on updated theoretical models, at \SI{18.1}{\mega\electronvolt} we take the same value quoted in~\cite{kr16}, $\Gamma_{\SI{18.1}{\mega\electronvolt}}=3.9 \times 10^{-3}$, while at \SI{17.6}{\mega\electronvolt} we use:

\begin{equation}
\Gamma_{\SI{17.6}{\mega\electronvolt}} = \frac{\Gamma^{ZM}_{\SI{17.6}{\mega\electronvolt}}}{\Gamma^{ZM}_{\SI{18.1}{\mega\electronvolt}}}\;\Gamma_{\SI{18.1} {\mega\electronvolt}} = 3.4 \times 10^{-3} \; ,
\end{equation}

where the $\Gamma^{ZM}_{Q}$ are computed from the Z-M model. For both values, we neglect the theoretical uncertainties.

The dependence of the X17 template on the mass of the X17 is introduced through vertical morphing \cite{BAAK201539}, a histogram interpolation technique performed on a bin-by-bin basis starting from a set of templates  for different X17 masses.
The relative IPC yields between the transition to ground state and first excited state were constrained based on BGO $\gamma$-ray spectra from the different proton energy bins:
\begin{itemize}
    \item $p_{\mathrm{IPC17.6}} = \mathcal{N}_{\mathrm{IPC17.6}}/(\mathcal{N}_{\mathrm{IPC14.6}} + \mathcal{N}_{\mathrm{IPC17.6}})$, expected to be  66.3 $\pm$ 1.7 \%,
    \item $p_{\mathrm{IPC17.9}} = \mathcal{N}_{\mathrm{IPC17.9}}/(\mathcal{N}_{\mathrm{IPC14.9}} + \mathcal{N}_{\mathrm{IPC17.9}})$, expected to be  48.2 $\pm$ 1.9 \% ,
    \item $p_{\mathrm{IPC18.1}} = \mathcal{N}_{\mathrm{IPC18.1}}/(\mathcal{N}_{\mathrm{IPC15.1}} + \mathcal{N}_{\mathrm{IPC18.1}})$, expected to be 43.0 $\pm$ 2.0 \%  ,
\end{itemize}

with the $\mathcal{N}_{\mathrm{IPC}}$ parameters the number of IPC events in the data set from their respective $\gamma$-ray line.
The uncertainty on the energy scale calibration due to the uncertainty on the value of the magnetic field, is accounted for by the nuisance parameter $\alpha_{field}$. Its effects are also introduced through vertical morphing.

The constraint terms, the energy scale and the ratio of the efficiencies are included in the likelihood through Gaussian terms.
The likelihood can be expressed as the product of three terms:
\begin{equation}
    \mathcal{L} = \mathcal{L}_{\mathrm{D}}\times\mathcal{L}_{\mathrm{S}} \times \mathcal{L}_{\mathrm{C}}
\end{equation}

with $\mathcal{L}_{\mathrm{D}}$ being the Poisson probability density function (PDF)  multiplied over the bins, $\mathcal{L}_{\mathrm{S}}$ being the term which accounts for the limited statistics of the Monte Carlo templates as defined in \cite{Dembinski2022} and $\mathcal{L}_{\mathrm{C}}$ the constraint term. For a given set of the parameters $\Omega$ = ($R_{17.6}$, $R_{18.1}$, $m_{\mathrm{X17}}$, $\mathcal{N}_{\mathrm{IPC}}$, $\mathcal{N}_{\mathrm{EPC}}$, $\mathcal{N}_{\mathrm{Fake}}$), of the Beeston-Barlow coefficients $\beta_i$ and of the nuisance parameters $\alpha_m$ = ($p_{\mathrm{IPC17.6}}$, $p_{\mathrm{IPC17.9}}$, $p_{\mathrm{IPC18.1}}$, $\alpha_{field}$, $k(m_{\mathrm{X17}})$):
\begin{eqnarray}
    \mathcal{L}_{\mathrm{D}}(\Omega, \alpha_m, \beta_i) &=& \prod_i \frac{f_i(\Omega, \alpha_m, \beta_i)^{D_i}e^{-f_i(\Omega, \alpha_m, \beta_i)}}{D_i!}\\
    \mathcal{L}_{\mathrm{S}}(\Omega, \alpha_m, \beta_i) &=& \prod_i \frac{(\beta_i\mu_{eff,i}(\Omega, \alpha_m))^{\mu_{eff,i}}e^{-\beta_i\mu_{eff,i}(\Omega, \alpha_m)}}{\mu_{eff,i}(\Omega, \alpha_m)!} \\
    \mathcal{L}_{\mathrm{C}}(\alpha_m) &=& \prod_m \frac{1}{\sqrt{2\pi}\sigma_{\alpha_m}}e^{-\frac{(\alpha_m - \alpha_{m,0})^2}{2 \sigma_{\alpha_m}^2}}\\
    f_i &=& \beta_i \sum_j \mathcal{N}_j a_{ij}
\end{eqnarray}

with $D_i$ the observed population of the $i$-th bin, $f_i$ its estimated population, obtained by summing the bins of the templates, running on index $j$, scaled by their yield $\mathcal{N}_j$. The statistical uncertainty $\sigma_{\beta_i}$ of the estimated bin population given by the limited statistics of the Monte Carlo template is used to compute the term $\mu_{eff,i}=1/\sigma_{\beta_i}^2$. The Beeston-Barlow parameter $\beta_i$ scales the bin population estimate to compensate such an effect.
A comparison of the fit model in the sideband data showed some limited discrepancy. The IPC templates, which were initially sampled with a statistics comparable to what was expected in the data, were scaled to $1/3$ of the best fit in the sidebands, effectively resulting in an inflation of their associated uncertainty.

The branching ratios are bound to be positive and the mass of the X17 is limited between \SI{16.5}{\mega\electronvolt/c^2} and \SI{17.1}{\mega\electronvolt/c^2}. The upper limit is given by the observation of the anomaly in the carbon de-excitation \cite{AtomkiCarbon}, which would be otherwise kinematically forbidden. The lower limit
is $2.5\sigma$ away from ATOMKI's best estimate.


\section{Branching ratio results}
\label{sec:branch}

\begin{figure*}
\centering
 \resizebox{1.99\columnwidth}{!}{%
 \includegraphics{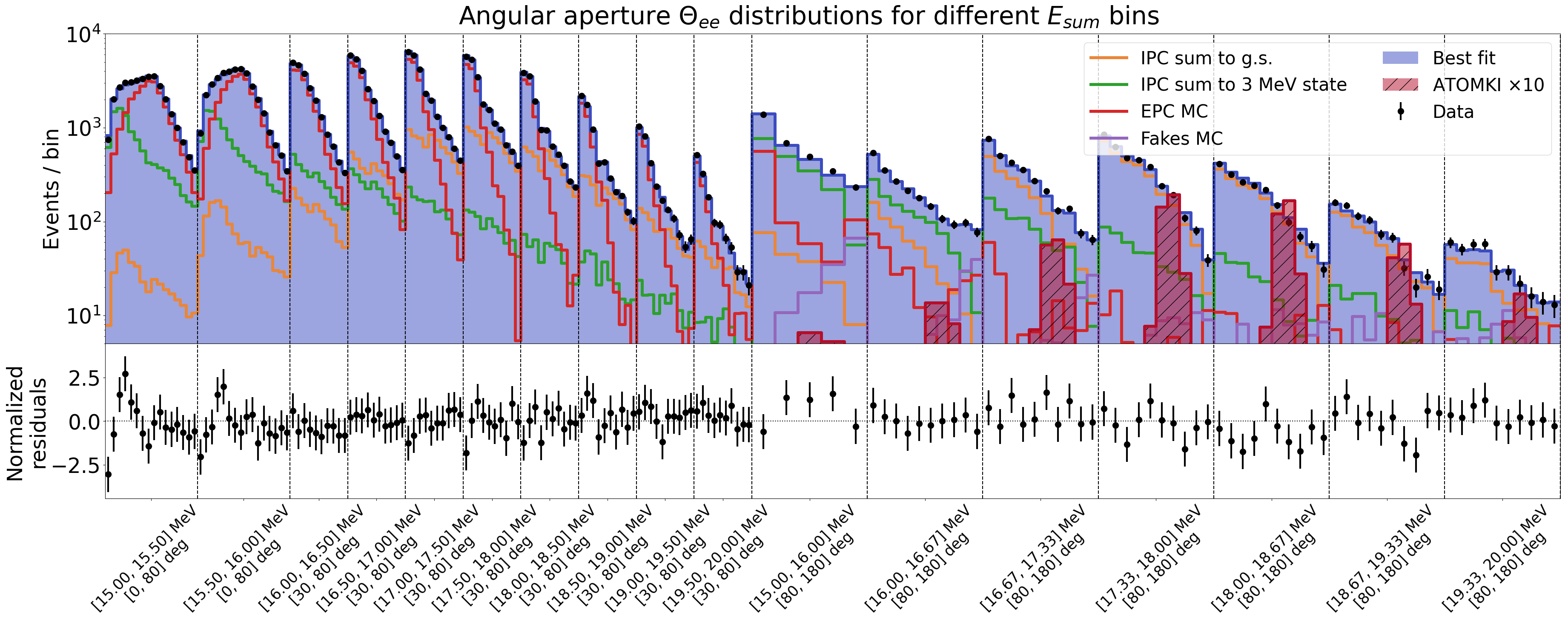}}
\caption{Top: comparison between the measured spectra of the angular opening \thetaee~(black dots) and the best fit (blue) for different $E_{sum}$ bins. The sum of the IPC populations from the de-excitation to ground state is shown in orange (IPC 18),  the sum of the IPC populations from the de-excitation to the first excited state is shown in green (IPC 15),  the sum of the EPC populations is shown in red and the population of fake events (see Sec.~\ref{subsec:eventsel}) is shown in purple. For comparison, the signal template is shown (hatched red) for a branching ratio 10 times larger than that measured at ATOMKI. Bottom: residuals normalized to the statistical uncertainty on the data points. }
\label{fig:fit2023}       
\end{figure*}


Figure~\ref{fig:fit2023} shows the result of the fit to the full 2023 data sample, reconstructed and selected as detailed above. The sum of the background contributions well describes the data. A goodness-of-fit test, based on the Poisson log-likelihood chi-square statistics defined in \cite{BAKER1984437}, returns a $p$-value of \SI{10.5}{\percent} by randomly generating a set of pseudo-experiments, indicating no significant deviation from the underlying model.

The best fit estimates \SI{12.6\pm0.9}{\percent} (\SI{45.8\pm1.3}{\percent}) of the IPC events coming from the \SI{18.1}{\mega\electronvolt} (\SI{17.6}{\mega\electronvolt}) resonance de-exciting to ground state and zero events from the \SI{17.9}{\mega\electronvolt} IPC. The best fit of the signal is obtained at $m_{\mathrm{X17}} = \SI{16.5}{\mega\electronvolt/c^2}$, with 10 events from the \SI{18.1}{\mega\electronvolt} resonance and with 0 from the \SI{17.6}{\mega\electronvolt} resonance.

We build confidence regions in the three-dimensional parameter space ($R_{17.6}$, $R_{18.1}$, $m_\mathrm{X17}$), adopting the Feldman-Cousins
approach~\cite{Feldman:1997qc}, with an ordering based on the profiled likelihood ratio:
\begin{equation}
    \lambda_p(R_{17.6}, R_{18.1}, m_\mathrm{X17}) = \frac{\mathcal{L}(R_{17.6}, R_{18.1}, m_\mathrm{X17}, \hat{\hat{\eta}})}{\mathcal{L}(\hat{R}_{17.6}, \hat{R}_{18.1}, \hat{m}_\mathrm{X17}, \hat{\eta})}
\end{equation}
with $\eta=(\mathcal{N}_{\mathrm{IPC}}, \mathcal{N}_{\mathrm{EPC}}, \mathcal{N}_{\mathrm{Fake}}, \alpha_m, \beta_i)$. The hats and double hats indicate the variables with respect to which the likelihood is maximized (compare Eq. 40.49 in \cite{ParticleDataGroup:2024cfk}). We conservatively quote the maximum extension of the confidence regions in the $\mathrm{R}_{Q}$'s as the upper limits on their values. 

\begin{figure*}
\centering
 \resizebox{!}{0.8\columnwidth}{%
 \includegraphics{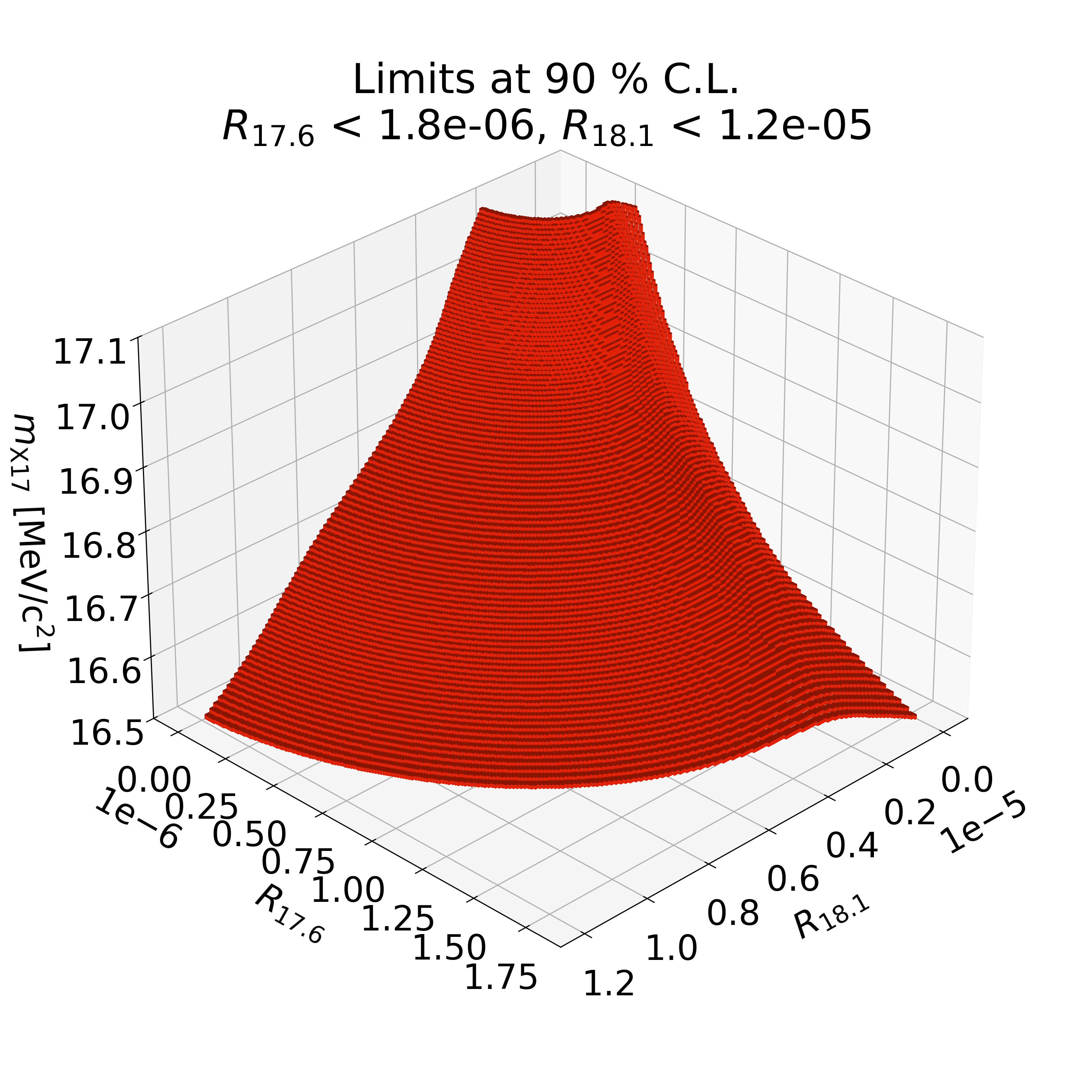}}
 \resizebox{!}{0.8\columnwidth}{%
 \includegraphics{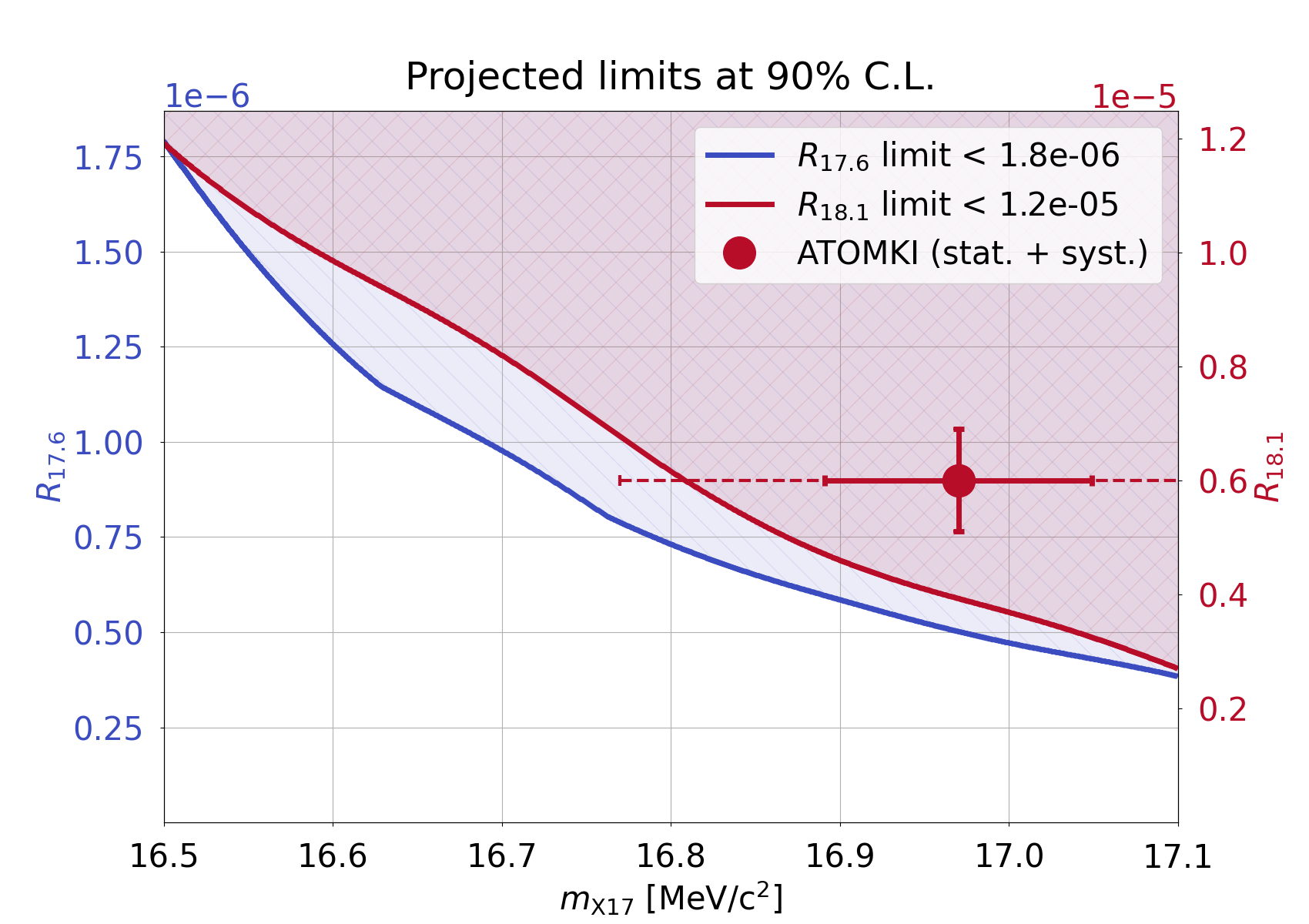} }
\caption{Left: 3D upper limits on $R_{17.6}$, $R_{18,1}$ and $m_\mathrm{X17}$ at \SI{90}{\percent} C.L.. Right: \SI{90}{\percent} C.L. limit projections of $R_{17.6}$ (blue) and $R_{18.1}$ (red) within the allowed mass range. The hatched area represents the excluded region. The red point represents the measured branching ratio at ATOMKI. The dashed error bar represents the systematic error on the X17 mass.}
\label{fig:FC}       
\end{figure*}

Fig.~\ref{fig:FC} shows the confidence regions for different values of confidence level (C.L.). The resulting 90\% C.L. region includes the null hypothesis, indicating no significant excess. We obtain $R_{18.1} < \SI{1.2e-5}{}$ and $R_{17.6} < \SI{1.8e-6}{}$, corresponding approximately to an upper limit of 230 and 200 respectively, on the total number of signal events.

We can also test some specific hypotheses on the mass and branching ratios, by randomly generating them in the pseudo-experiments according to Gaussian distributions with mean and sigma derived from measurements or theoretical models with their uncertainties. In this case, a Gaussian constraint on the mass is also included in the likelihood ratio, which is defined as:
\begin{equation}
    \lambda_{pH} = \frac{\mathcal{L}( \overline{R_{17.6}}, \overline{R_{18.1}}, \overline{m_\mathrm{X17}}, \hat{\hat{\eta}}) \, G( \overline{m_\mathrm{X17}})}{\mathcal{L}(\hat{R}_{17.6}, \hat{R}_{18.1}, \hat{m}_\mathrm{X17}, \hat{\eta}) \, G(\hat{m}_\mathrm{X17})} \; ,
\end{equation}

where the bar indicates the mean value of the assumed Gaussian distribution. This likelihood ratio is calculated for any pseudo-experiment, and the $p$-value for the hypothesis under study is defined as the fraction of pseudo-experiments producing a smaller $\lambda_{pH}$ value than the one obtained on data. 
We consider here two hypotheses:
\begin{itemize}
\item ATOMKI hypothesis: $R_{17.6}$ is fixed to zero, according to the claim that no evidence of X17 production was observed from the 17.6~MeV state decay. The X17 mass and $R_{18.1}$ are generated according to Gaussian distributions, with means and widths taken from a combination of Atomki's results on all three nuclei \cite{kr16, kr141,kr17,Krasznahorkay:2019lyl,AtomkiCarbon}: $m_{\mathrm{X17}} = \SI{16.97\pm 0.22}{\mega\electronvolt/c^2}$, $\mathrm{R}_{18.1}=\SI{6\pm1e-6}{}$.
\item Feng {\it et al.} hypothesis: The X17 is supposedly produced from both $^8$Be$^{*}$(17.6) and $^8$Be$^{*}$(18.1). The X17 mass and $R_{18.1}$ are generated as in the previous hypothesis. Starting from the generated value of $R_{18.1}$, $R_{17.6}$ is fixed with a scaling by a kinematical suppression factor of 0.46, as prescribed in~\cite{Feng:2016jff}.
\end{itemize}
$p$-values of \SI{6.2}{\percent} ($1.5\sigma$) and \SI{1.80}{\percent} ($2.1\sigma$) are obtained for these two hypotheses respectively. 

\section{Conclusions}
\label{sec:concl}
A search for the hypothetical X17 particle suggested to explain an anomaly first observed at an experiment in ATOMKI was conducted with the MEG II apparatus at PSI.   The  \ee~pair production in the nuclear transitions of  both 17.6 MeV and 18.1 MeV $^8$Be  excited states produced in  the nuclear interaction of protons with $^7$Li was studied. In our dataset, $^8\mathrm{Be}^*(18.1)\rightarrow {}^8\mathrm{Be}+\mathrm{e^+e^-}$ represents 21.6(25)\% of all IPC transitions to ground state. No significant evidence of the X17 particle was found. Upper limits at 90\% C.L. were set on the X17 production branching ratio with respect to $\gamma$-ray emission, $R_{18.1} < \SI{1.2e-5}{}$ and $R_{17.6} < \SI{1.8e-6}{}$, for X17 masses between \SI{16.5}{\mega\electronvolt/c^2} and \SI{17.1}{\mega\electronvolt/c^2}. The Atomki X17 hypothesis was tested and a 6.2\% (1.5$\sigma$) $p$-value was obtained. Improved sensitivity to the particle production can be achieved with higher statistics to be collected at the 1030~keV resonance.

\section{Acknowledgements}
\label{sec:acknow}
We are grateful for the support and cooperation provided by PSI as the host laboratory and to the technical and engineering staff of our institutes. This work is supported by INFN (Italy), the Italian Ministry of University and Research (MIUR) grant 2022ENJMRS, Marie Skłodowska-Curie Innovative Training Networks grant 858199 (Horizon 2020), JSPS KAKENHI numbers JP26000004,
20H00154,21H04991,21H00065,22K21350 and JSPS Core-to-Core
Program, A. Advanced Research Networks JPJSCCA20180004 (Japan), the Leverhulme Trust, LIP-2021-01 (UK).

%

\bibliography{x17biblio}

\begin{thebibliography}{10}
\providecommand{\url}[1]{{#1}}
\providecommand{\urlprefix}{URL }
\expandafter\ifx\csname urlstyle\endcsname\relax
  \providecommand{\doi}[1]{DOI \discretionary{}{}{}#1}\else
  \providecommand{\doi}{DOI \discretionary{}{}{}\begingroup \urlstyle{rm}\Url}\fi

\bibitem{kr16}
A.J. Krasznahorkay, et~al., Phys. Rev. Lett. \textbf{116}, 042501 (2016).
\newblock \urlprefix\url{https://link.aps.org/doi/10.1103/PhysRevLett.116.042501}

\bibitem{TILLEY2004155}
D.~Tilley, et~al., Nuclear Physics A \textbf{745}(3), 155 (2004).
\newblock \urlprefix\url{https://www.sciencedirect.com/science/article/pii/S0375947404010267}

\bibitem{SCHLUTER1981327}
P.~Schlüter, G.~Soff, W.~Greiner, Physics Reports \textbf{75}(6), 327 (1981).
\newblock \urlprefix\url{https://www.sciencedirect.com/science/article/pii/0370157381901666}

\bibitem{PhysRev.76.678}
M.E. Rose, Phys. Rev. \textbf{76}, 678 (1949).
\newblock \doi{10.1103/PhysRev.76.678}.
\newblock \urlprefix\url{https://link.aps.org/doi/10.1103/PhysRev.76.678}

\bibitem{kr141}
{Krasznahorkay, A.J.}, et~al., EPJ Web Conf. \textbf{142}, 01019 (2017).
\newblock \urlprefix\url{https://doi.org/10.1051/epjconf/201714201019}

\bibitem{kr17}
A.~Krasznahorkay, et~al., in \emph{{Proceedings, 55th International Winter Meeting on Nuclear Physics, January 23-27, 2017}} (2017)

\bibitem{Krasznahorkay:2019lyl}
A.J. Krasznahorkay, et~al.,   (2019).
\newblock \urlprefix\url{https://arxiv.org/abs/1910.10459}

\bibitem{AtomkiCarbon}
A.J. Krasznahorkay, et~al., Phys. Rev. C \textbf{106}, L061601 (2022).
\newblock \urlprefix\url{https://link.aps.org/doi/10.1103/PhysRevC.106.L061601}

\bibitem{Alves:2023ree}
D.S.M. Alves, et~al., Eur. Phys. J. C \textbf{83}(3), 230 (2023).
\newblock \urlprefix\url{https://link.aps.org/doi/10.1140/epjc/s10052-023-11271-x}

\bibitem{Anh2024}
T.T. Anh, et~al., Universe \textbf{10}(4) (2024).
\newblock \urlprefix\url{https://www.mdpi.com/2218-1997/10/4/168}

\bibitem{Abraamyan2023}
K.U. Abraamyan, et~al.,   (2023).
\newblock \urlprefix\url{https://doi.org/10.48550/arXiv.2311.18632}

\bibitem{MEGII:2023fog}
K.~Afanaciev, et~al., Eur. Phys. J. C \textbf{84}(2), 190 (2024).
\newblock \urlprefix\url{https://doi.org/10.1140/epjc/s10052-024-12415-3}

\bibitem{MEG:2011rgj}
J.~Adam, et~al., Nucl. Instrum. Meth. A \textbf{641}, 19 (2011).
\newblock \urlprefix\url{https://doi.org/10.1016/j.nima.2011.03.048}

\bibitem{DCHMEGperformances}
A.M. Baldini, et~al., The European Physical Journal C \textbf{84}(5), 473 (2024).
\newblock \urlprefix\url{https://doi.org/10.1140/epjc/s10052-024-12711-y}

\bibitem{Baldini_2018}
A.~Baldini, et~al., Journal of Instrumentation \textbf{13}(06), P06018 (2018).
\newblock \urlprefix\url{https://dx.doi.org/10.1088/1748-0221/13/06/P06018}

\bibitem{ref:MEGIIdesign}
A.M. Baldini, et~al., The European Physical Journal C \textbf{78}(5), 380 (2018).
\newblock \urlprefix\url{https://doi.org/10.1140/epjc/s10052-018-5845-6}

\bibitem{ADAM201119}
J.~Adam, et~al., Nuclear Instruments and Methods in Physics Research Section A: Accelerators, Spectrometers, Detectors and Associated Equipment \textbf{641}(1), 19 (2011).
\newblock \urlprefix\url{https://www.sciencedirect.com/science/article/pii/S0168900211006875}

\bibitem{Viviani:2021stx}
M.~Viviani, et~al., Phys. Rev. C \textbf{105}(1), 014001 (2022).
\newblock \urlprefix\url{https://doi.org/10.1103/PhysRevC.105.014001}

\bibitem{FRANCESCONI2023167542}
M.~Francesconi, et~al., Nuclear Instruments and Methods in Physics Research Section A: Accelerators, Spectrometers, Detectors and Associated Equipment \textbf{1045}, 167542 (2023).
\newblock \urlprefix\url{https://www.sciencedirect.com/science/article/pii/S0168900222008348}

\bibitem{Agostinelli2003250}
S.~Agostinelli, et~al., Nucl. Instr. Meth. \textbf{A 506}(3), 250  (2003).
\newblock \doi{DOI: 10.1016/S0168-9002(03)01368-8}.
\newblock \urlprefix\url{http://www.sciencedirect.com/science/article/pii/S0168900203013688}

\bibitem{Zhang:2017zap}
X.~Zhang, G.A. Miller, Physics Letters B \textbf{773}, 159 (2017).
\newblock \urlprefix\url{https://www.sciencedirect.com/science/article/pii/S0370269317306342}

\bibitem{Blatt:1952ije}
J.M. Blatt, V.F. Weisskopf, \emph{{Theoretical nuclear physics}} (Springer, New York, 1952).
\newblock \doi{10.1007/978-1-4612-9959-2}

\bibitem{Dembinski2022}
H.~Dembinski, A.~Abdelmotteleb, The European Physical Journal C \textbf{82}(11), 1043 (2022).
\newblock \urlprefix\url{https://doi.org/10.1140/epjc/s10052-022-11019-z}

\bibitem{BAAK201539}
M.~Baak, et~al., Nuclear Instruments and Methods in Physics Research Section A: Accelerators, Spectrometers, Detectors and Associated Equipment \textbf{771}, 39 (2015).
\newblock \urlprefix\url{https://www.sciencedirect.com/science/article/pii/S0168900214011814}

\bibitem{BAKER1984437}
S.~Baker, R.D. Cousins, Nuclear Instruments and Methods in Physics Research \textbf{221}(2), 437 (1984).
\newblock \urlprefix\url{https://www.sciencedirect.com/science/article/pii/0167508784900164}

\bibitem{Feldman:1997qc}
G.J. Feldman, R.D. Cousins, Phys. Rev. D \textbf{57}, 3873 (1998).
\newblock \urlprefix\url{https://doi.org/10.1103/PhysRevD.57.3873}

\bibitem{ParticleDataGroup:2024cfk}
S.~Navas, et~al., Phys. Rev. D \textbf{110}(3), 030001 (2024).
\newblock \urlprefix\url{https://doi.org/10.1103/PhysRevD.110.030001}

\bibitem{Feng:2016jff}
J.L. Feng, et~al., Phys. Rev. Lett. \textbf{117}(7), 071803 (2016).
\newblock \urlprefix\url{https://doi.org/10.1103/PhysRevLett.117.071803}

\end{thebibliography}
\bibliographystyle{spphys}


\end{document}